\documentclass[conference]{IEEEtran}

\usepackage{graphicx}
\graphicspath{ {./images/} }
\usepackage[backend=biber,style=numeric,sorting=none]{biblatex}
\usepackage{csquotes}
\usepackage{siunitx}
\usepackage{subcaption}
\usepackage{hyperref}
\usepackage{amssymb}
\usepackage{amsmath}
\usepackage[svgnames]{xcolor}
\hypersetup{
    colorlinks=true,
    citecolor=Grey,
    linkcolor=RoyalBlue,      
    urlcolor=cyan
}

\usepackage{dblfloatfix}
\usepackage{array}

\ifCLASSINFOpdf
\else
\fi
\begin{document}
%
\title{Robustness Evaluation of\\Stacked Generative Adversarial Networks using\\Metamorphic Testing}

\author{
\IEEEauthorblockN{Hyejin Park\IEEEauthorrefmark{1}, Taaha Waseem\IEEEauthorrefmark{1}, Wen Qi Teo\IEEEauthorrefmark{1}, Ying Hwei Low\IEEEauthorrefmark{1}, Mei Kuan Lim\IEEEauthorrefmark{1}, Chun Yong Chong\IEEEauthorrefmark{1}}
\IEEEauthorblockA{\IEEEauthorrefmark{1}School of Information Technology,\\ 
Monash University Malaysia, 47500, Bandar Sunway, Selangor, Malaysia\\
\{hpar35, twas0001, wteo0006, ylow0008\}@student.monash.edu  \{lim.meikuan, chong.chunyong\}@monash.edu}
}



%


\maketitle

\begin{abstract}
Synthesising photo-realistic images from natural language is one of the challenging problems in computer vision. Over the past decade, a number of approaches have been proposed, of which the improved Stacked Generative Adversarial Network (StackGAN-v2) has proven capable of generating high resolution images that reflect the details specified in the input text descriptions. In this paper, we aim to assess the robustness and fault-tolerance capability of the StackGAN-v2 model by introducing variations in the training data. However, due to the working principle of Generative Adversarial Network (GAN), it is difficult to predict the output of the model when the training data are modified. Hence, in this work, we adopt Metamorphic Testing technique to evaluate the robustness of the model with a variety of unexpected training dataset. As such, we first implement StackGAN-v2 algorithm and test the pre-trained model provided by the original authors to establish a ground truth for our experiments. We then identify a metamorphic relation, from which test cases are generated. Further, metamorphic relations were derived successively based on the observations of prior test results. Finally, we synthesise the results from our experiment of all the metamorphic relations and found that StackGAN-v2 algorithm is susceptible to input images with obtrusive objects, even if it overlaps with the main object minimally, which was not reported by the authors and users of StackGAN-v2 model. The proposed metamorphic relations can be applied to other text-to-image synthesis models to not only verify the robustness but also to help researchers understand and interpret the results made by the machine learning models. 
\end{abstract}

\begin{IEEEkeywords}
Metamorphic testing, Stacked Generative Adversarial Network, metamorphic relations, robustness testing.
\end{IEEEkeywords}


%
\IEEEpeerreviewmaketitle

\section{Introduction}
Text-to-Image synthesis has become an important research topic in the area of artificial intelligence \cite{pan2019gansurvey}. The introduction of Generative Adversarial networks (GANs) by Goodfellow et al. \cite{goodfellow2014gan} in 2014 has enabled rapid improvement in state-of-the-art methods on text-to-image synthesis. Built upon GANs, the Stacked Generative Adversarial Network (StackGAN) proposed by Zhang et al. in 2016 \cite{zhang2016stackgan} has enhanced the stability of GAN’s training process and the output resolution of the samples generated. The authors of the StackGAN have proposed the improved StackGAN-v2 \cite{zhang2018stackgan++} in the following year, further improving the training model with a series of multi-scale image distributions. 

In most existing image synthesis research works, the training images used are of the best quality with a clear, high resolution photo of a target, which might be hard to collect and produce in practical scenarios. In most of the use cases, the images used for machine learning may have deformation. Noise distortion can arise with multiple objects in the image and watermark added. The issue of motion blur can also occur due to a moving camera or moving target objects. In certain scenarios, the image may not fully capture the target objects but is represented as a cropped part of target. Often, the performance of the generative models is susceptible to small image transformations \cite{samuel2016imagequality} in contrast to robust human visual systems \cite{samuel2017humancomparison}. Therefore, in this paper, we focus on robustness evaluation of StackGAN-v2 specifically on image transformation, by introducing foreign objects into the training image dataset and investigate the quality of the image generated from the altered images. These images are generated based on several metamorphic relations that we defined in this study and Inception Score (IS) will be used to assess and compare the results. 

With these in mind, the following research questions need to be addressed:
\begin{itemize}
    \item How can metamorphic relations between the characteristics of the model and its robustness be identified?
    \item How do variations in training data affect the IS of the synthesised images?
\end{itemize}

This paper aims to develop a means of generating metamorphic relations to aid the evaluation and interpretation of image synthesis models. Specifically, the objectives of this paper includes:
\begin{itemize}
    \item To propose metamorphic relations that are capable of examining the relations between the characteristics of the StackGAN-v2 model and its robustness;
    \item To identify the effects of variations in training data on the generated model.
\end{itemize}


We use Inception Score (IS) \cite{barratt2018note} as the metric to evaluate the quality of the images generated from StackGAN-v2 model. IS is a widely used metric to evaluate the quality of an image generative model. 
\[ \mathrm{IS}(G) = \exp (\mathbb{E}_{\mathrm{x}\sim p_{g}} D_{KL}(p(y|\mathrm{x})\|p(y))) \]
where \(\mathrm{x}\sim p_{g}\) denotes a generated sample \(\mathrm{x}\) from distribution of images \(p_{g}\), and \(y\) is the label classified by inception model. \(D_{KL}(p(y|\mathrm{x})\|p(y))\) indicates $KL$ divergence between the conditional class distribution \(p(y|\mathrm{x})\) and the marginal class distribution \(p(y)\). Therefore, if a generative model is capable of generating a high diversity of images, large $KL$ divergence is expected, leading to a high IS.

While IS is able to evaluate the model quantitatively, it is difficult to capture the correctness of the generated images with respect to the intended output from a given input text, given the black-box nature of the StackGAN-v2 model.
As such, we propose a metamorphic testing approach to better understand the synthesized image, given a set of input data. 
Instead of identifying every metamorphic relation (MR) prior to conducting the experiments, we employ investigative steps to generate MR based on the analysis of test case results from the previous metamorphic relation consecutively, to verify and emphasise the findings from the previous test case.


Using our motivating example that it is extremely difficult to collectivise perfect training images without any noise, we first propose a metamorphic relation to evaluate the robustness of StackGAN-v2 algorithm when there are presences of a foreign object in the training data.
As experiments are conducted, the unknown behaviour of StackGAN-v2 is revealed, from which new metamorphic relations are further identified. Details of the experiments on these metamorphic relations are given in Section \ref{experiments}. 

\begin{figure}[t]
\includegraphics[width=0.25\textwidth]{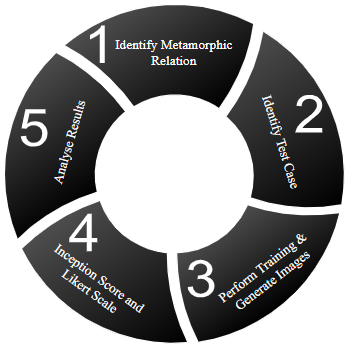}
\centering
\caption{Overall workflow for this study.}
\label{fig:methodology_flowchart}
\end{figure}

The main discovery and contribution of this study are that metamorphic testing works effectively in an iterative process model. While most of the metamorphic testing project had defined the relations prior based on the expected outcome, like the testing done in the work by \cite{wang2019metamorphic, zhou2019driverlesscar, zhang2018metamorphicautonomousdriving}, This paper shows that metamorphic testing can follow an iterative or cyclic process flow, similarly to how we have conducted our study as summarised in Figure \ref{fig:methodology_flowchart}.
This is a significant finding in the field of software testing, especially in applications where the algorithm is too complex for researchers to predict the outcome of a given input.

\section{Background}

\subsection{Stacked Generative Adversarial Networks} 
Stacked Generative Adversarial Network (StackGAN) \cite{zhang2016stackgan} is a framework designed based on Generative Adversarial Network (GAN) \cite{goodfellow2014gan} to synthesise high-quality images with photo-realistic details reflecting the context of the given text description. StackGAN is designed as a two-stage GAN to address the instability issue of GANs’ training process upon high resolution image generation. Stage-I GAN first generates low resolution images by sketching the primitive shape and basic colours of the target described in the given text input. Then, Stage-II GAN improves the semantics of the Stage-I result images by conditioning with the text input again to capture the omitted details. 

An improved version of StackGAN, called the StackGAN-v2 \cite{zhang2018stackgan++}, was proposed to further refine and stabilise the earlier model architecture by arranging multi-stages of GAN in a tree-like structure. The root of the tree represents the input dataset, and the deepest level of the branches provides the final output of generated samples, where the most semantic and photo-realistic images are drawn. The intermediate GANs gradually improve resolution of the sample to be generated at each branch level. In this manner, the generative model of GAN at each branch learns to capture image distributions, and the feedback from one model can improve the learning process of other generative models at different branches. The discriminative models take the original datasets and samples from generative models with respective conditioning variables to train for classification.

\subsection{Metamorphic Testing}
Metamorphic testing \cite{10.1145/3143561} is an approach widely used for testing machine learning systems to handle the oracle problem as test oracle may not always exist \cite{wang2019metamorphic}. 

In metamorphic testing, new test cases are generated based on the set of input-output pairs from the previous test cases to better understand the relation among the series of inputs and their corresponding outputs, with what is known as metamorphic relation (MR). The implementation of metamorphic testing can be broken down into three steps \cite{segura2016surveyonmt, Zhou2019Meta}. 

\begin{enumerate}
    \item Identify the characteristics of the program under test and represent them in the form of relations among a set of test inputs and their corresponding expected outputs. 
    \item Generate source test cases using traditional test case selection techniques. The source test cases are used to serve as seeds to generate the follow-up test cases using identified metamorphic relation.
    \item  Execute the test cases to verify whether the outputs of the source test case and follow-up test cases have satisfied or violated the metamorphic relation they were built on.
\end{enumerate}

In this paper, we apply metamorphic testing to show how the unexpected image dataset can impact the performance of the StackGAN-v2 model. 

\section{Related Work}
Most of the GAN models are susceptible to errors caused by the discrepancies in training data. Not only is there a lack of test oracles to ensure efficiency and correctness of machine learning models, but there is also a lack of metrics to evaluate model correctness and image quality. Metamorphic testing \cite{zhou2019driverlesscar} is chosen in this study to augment existing testing strategies and to identify unknown limitations of the current StackGAN-v2 model.

Metamorphic testing has shown tremendous success in evaluating the robustness of self-driving or autonomous driving algorithms because the testing of these algorithms lacks diversity in test cases. This is largely due to the complexity involved in generating driving scenes which are known to impair driving abilities \cite{zhang2018metamorphicautonomousdriving}. The problem was resolved with the employment of a GAN-based technique to synthesise driving scenes with varying weather conditions, and the expected steering angle signals were then predicted based on the metamorphic relations \cite{zhang2018metamorphicautonomousdriving}. Metamorphic testing was used to detect violations of the metamorphic relations through prompt analysis of the outputs to evaluate robustness of the autonomous driving system.

Furthermore, existing studies found that testing on object detection algorithms experienced poor test coverage due to lack of diversity and realness in test inputs \cite{wang2019metamorphic}. MetaOD, a combination of metamorphic testing with image processing to add different objects into background images was introduced to generate modified test inputs that look realistic \cite{wang2019metamorphic}. MetaOD does so by checking the consistency of object detection results between synthetic images and the corresponding backgrounds by employing metamorphic relations and evaluating the extent to which they are upheld \cite{wang2019metamorphic}.

However, the success of metamorphic testing heavily relies on the quality of metamorphic relations, which are not straightforward and difficult to identify. Metamorphic relation pattern (MRP) is one of the methods that can be utilised to identify metamorphic relations \cite{Zhou2020MetamorphicRF}. Examples of metamorphic relation patterns include input equivalence and shuffling \cite{Segura2019MetamorphicRP}. Apart from this, existing studies have also utilised symmetry as a metamorphic relation pattern which assumes that the system would appear the same under different viewpoints \cite{Zhou2020MetamorphicRF}.

This study is motivated by the ability of metamorphic testing to exploit metamorphic relations in evaluating the robustness of an algorithm \cite{wang2019metamorphic} \cite{zhou2019driverlesscar} \cite{zhang2018metamorphicautonomousdriving}. Thus, in the context of our work, the first step is to identify the potential metamorphic relations between the training dataset and the photo-realistic images generated by the model. Test data is then altered and manipulated correspond to the metamorphic relations, in order to identify the unknown limitations or faults of the StackGAN-v2 model, that are otherwise hidden from the researchers and users. Ultimately, the findings of our research can be used to evaluate the robustness of StackGAN algorithm.

\section{Methods} \label{methods}
We setup an experimental test bed using \textit{GeForce RTX 2080 Ti} with \textit{CUDA 10.0} to allow the execution of the study objectives. Further details of our implementation can be found at our GitHub page\footnote{https://github.com/FIT4003StackGAN/FIT4003-StackGAN-v2-Metamorphic-Testing}.

In general, we employ investigative steps to identify the metamorphic relations in the StackGAN-v2 model by manipulating the training image dataset. This is necessary and extremely crucial to our study as we cannot predict the output of the model when given a set of unexpected inputs, and thus are unable to identify more metamorphic relations without performing some form of initial analysis. 

We introduce noise to the entirety of the training image dataset by adding foreign object bodies that tend to be absent from the original dataset, such as an addition of bird or trees. The modified images are inspected to ensure minimal obstruction of the focal objects, which in this study are birds. Training would then be performed on the modified images, which requires approximately 33 hours to run on 600 epochs (as per recommended in StackGAN-v2). Following this, we generate the images for the model which are then analysed using two different evaluation metrics, namely the IS and the Likert Scale \cite{joshi2015likert}. Finally, we decipher the results of the test case, and from there, design and identify new metamorphic relations and the cycle continues. The detailed implementation of our experiments are discussed in Section \ref{experiments}.

\section{Experiments} \label{experiments}

\subsection{Data Collection}
We evaluate StackGAN-v2 on the CUB \cite{WahCUB_200_2011} datasets which contain 200 bird species with 11,788 images, as was utilised in the original StackGAN-v2 paper. We utilised the image list for class-disjoint training and test sets as specified by the original StackGAN-v2 paper without any modifications, to reproduce the results as closely as possible. This is done so that the results from the unmodified datasets may act as the ground truth for our experiments.

\subsection{Evaluation Metrics}
It is important that we apply the same metrics used to get an unbiased comparison, and so we utilise the IS model fine-tuned by StackGAN-v2 as our quantitative evaluation metric. The higher the IS for a set of images, the better the quality and diversity of the images. All reported ISs in this study will be utilising the best result of each test case.

\begin{table}[t]
\caption{Comparison of ISs and Likert Scale Evaluation between Test Cases}
\label{table:1}
\tiny
\centering
\sisetup{
 table-number-alignment = center,
 table-figures-integer = 1
}
\begin{tabular}{ 
 |S|
 *{3}{@{\hspace{0.5\tabcolsep}}S[
  separate-uncertainty,
  table-figures-uncertainty = 1
 ]@{\hspace{0.5\tabcolsep}}|}
 } 
 \hline
 {Test Cases} & {IS} &
 {Semantic Likert Scale} & 
 {Realistic Likert Scale} \\ 
 \hline
 
 {StackGAN-v2 paper} & 4.05(5) & {-} & {-} \\ 
 {Pre-Trained} & 4.08(5) & {-} & {-} \\
 {Self-Trained} & 4.16(3) & 2.37(88) & 2.59(111) \\
 {\textbf{TC\textsubscript{01}}} & 3.50(4) & 1.60(85) & 1.73(100) \\
 {\textbf{TC\textsubscript{02}}} & 3.84(4) & 1.48(62) & 1.76(99) \\
 {\textbf{TC\textsubscript{03}}} & 3.99(5) & 1.75(88) & 1.84(110) \\
 {\textbf{TC\textsubscript{04}}} & 3.96(6) & 1.18(38) & 1.15(42) \\
 {\textbf{TC\textsubscript{05}}} & 3.92(4) & 1.98(80) & 2.25(101) \\
 {\textbf{TC\textsubscript{06}}} & 4.06(2) & 1.89(97) & 1.90(98) \\
 {\textbf{TC\textsubscript{07}}} & 3.88(5) & 1.91(91) & 1.77(107) \\
 {\textbf{TC\textsubscript{08}}} & 3.83(3) & 1.68(70) & 1.68(92) \\
 \hline
\end{tabular}
\end{table}

While studies have shown that the IS correlates well with human judgement, it is our concern that the usage of a fine-tuned model provided by StackGAN-v2 may introduce some form of partiality in our experiments. Thus, we perform an additional analysis on the generated images with the highest IS for each test case. The set of highest quality (IS) images would be evaluated using the Likert Scale -- a form of evaluation metric to quantify human thoughts and perception \cite{joshi2015likert}. This quantitative analysis is conducted amongst the co-authors, using a five-point scale with 1 representing very poor, and 5 representing excellent. For all test cases, we compare the generated images with the training image datasets in terms of semantics and photo-realism. A total of 40 generated images are chosen for the Likert Scale evaluation of the White breasted Kingfisher species. The compilation of the details of the Likert Scale evaluation will be elaborated towards the end of the Section \ref{results}.

Finally, much of this study involves qualitative analysis, where we manually review the images by eye and noting interesting details for each result. Explanations of our observations are elaborated in Section \ref{results}.

\subsection{Preliminaries} \label{preliminaries}
Metamorphic relations are discovered and revised as we progress through our experiments and reveal more behaviour of the StackGAN-v2 model. We were able to identify a metamorphic relation prior to conducting the experiments which are built on sound reasoning. The following metamorphic relation continue to be relevant throughout the course of this study and is proposed as follows:
\begin{displayquote}
\textbf{MR\textsubscript{01}} Introduction of a minimally obtrusive object consistently in all training image datasets should not drastically affect the IS.
\end{displayquote}
\textbf{MR\textsubscript{01}} assumes that the secondary object introduced into the training image should not severely affect the results of the IS, especially when the object has minimal obscuration on the main focus of the image, which is a single bird. This metamorphic relation was built on observations of the unmodified training datasets, which were found to contain watermarks in more than one instance. Thus, we find it reasonable to assume the added object should not affect the results by a significant factor. 

Prior to executing any test cases, we tested the pre-trained model provided by StackGAN-v2 and performed training on the original dataset to establish a stable ground truth. In the following sections, the term \textit{self-trained} will be used to denote the model trained from unmodified dataset. We found that we were able to closely replicate the IS provided in the StackGAN-v2 paper, and the generated images were relatively realistic in nature. The compilations of the ISs are presented in Table \ref{table:1}. In addition to the birds model, we have also initialised the church model for investigative purposes.

\begin{figure}[t]
\includegraphics[width=0.48\textwidth]{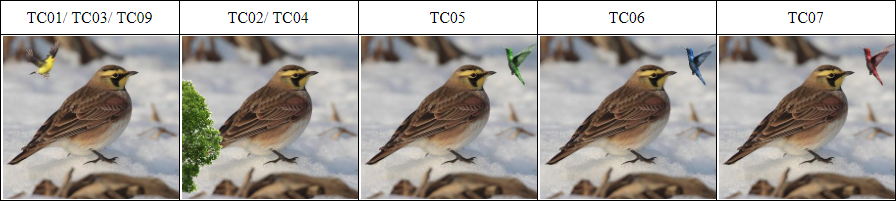}
\centering
\caption{Sample of the modified training images in different test cases.}
\label{fig:sample_modified_img}
\end{figure}

Having established the fundamental metamorphic relation, we began exploring objects of interest to insert into the training images. These objects need to be causing minimal obstruction to the principal object, the bird, and still be of theoretical interest to our study. The introduced objects were also scaled to a manageable size such that it would be smaller in size than the principal object. Manual validation was conducted to ensure the introduced objects result in minimal obstruction of the view of the principal object. Samples of the modified images are presented in Figure \ref{fig:sample_modified_img}, where the labels \textit{TC01-TC08} correspond to the test cases generated from different \textbf{MRs}, which will be discussed in the subsequent sections. Furthermore, a summary that presents the associated description and test cases of all the MRs that were covered in this paper are available on Table \ref{table:2}.

\subsection{Test Cases and Results} \label{results}

\begin{figure}[t]
\includegraphics[width=0.35\textwidth]{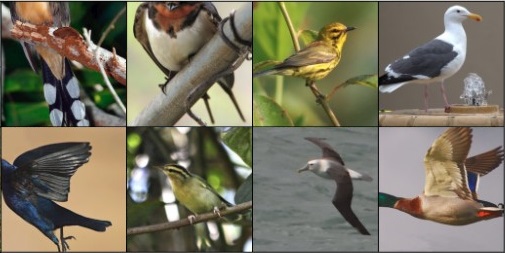}
\centering
\caption{The real samples images for self-trained model. Real samples are the processed training image through usage of bounding boxes to identify the focal birds. }
\label{fig:bounding_box_issue}
\end{figure}

\begin{table*}[h!]
\caption{Summarised Table of MRs}
\label{table:2}
\centering
\sisetup{
 table-number-alignment = center,
 table-figures-integer = 1
}
  \begin{tabular}{|m{0.04\linewidth}| @{}c@{}|}
    \hline
    MRs & Information \\
    \hline
   {MR\textsubscript{01}} &
   \begin{tabular}{m{0.1\linewidth}| @{}c@{}}
      {Description} &
      \begin{tabular}{m{0.8\linewidth}}
      {Introduction of a minimally obtrusive object consistently in all training image dataset should not drastically affect the IS.}
      \end{tabular}
      \\\hline
      {Causal Relation} &
       \begin{tabular}{m{0.8\linewidth}}
      {-}
      \end{tabular}
      \\\hline
      {Test Cases} & 
      \begin{tabular}{m{0.03\linewidth}|@{}c@{}}
        {TC\textsubscript{01}} & 
        \begin{tabular}{m{0.08\linewidth}|m{0.64\linewidth}}
             {Modification} & {Introduction of one bird to 100\% of the training images.} \\\hline
             {Variation} & {10 different birds chosen at random for each image.}\\
        \end{tabular}
        \\
        \end{tabular} \\
    \end{tabular} \\\hline
    {MR\textsubscript{02}} &
   \begin{tabular}{m{0.1\linewidth}| @{}c@{}}
      {Description} &
      \begin{tabular}{m{0.8\linewidth}}
      {Introduction of a minimally obtrusive object consistently in all training image dataset should result in a grey-tinted effect and impact the IS in a similar manner, regardless of the type of object introduced.}
      \end{tabular}
      \\\hline
      {Causal Relation} &
       \begin{tabular}{m{0.8\linewidth}}
      {MR\textsubscript{01}}
      \end{tabular}
      \\\hline
      {Test Cases} & 
      \begin{tabular}{m{0.03\linewidth}|@{}c@{}}
        {TC\textsubscript{02}} & 
        \begin{tabular}{m{0.08\linewidth}|m{0.64\linewidth}}
             {Modification} & {Introduction of one tree to 100\% of the training images.} \\\hline
             {Variation} & {10 different trees chosen at random for each image.}\\
        \end{tabular}
        \\
        \end{tabular} \\
    \end{tabular} \\\hline
    {MR\textsubscript{03}} &
   \begin{tabular}{m{0.1\linewidth}| @{}c@{}}
      {Description} &
      \begin{tabular}{m{0.8\linewidth}}
      {Introduction of a minimally obtrusive object in only a selected portion of the training image dataset should result in a diminished effect of the grey-tinted effect and a higher IS than the model in which all training image dataset were modified.}
      \end{tabular}
      \\\hline
      {Causal Relation} &
       \begin{tabular}{m{0.8\linewidth}}
      {MR\textsubscript{01}}
      \end{tabular}
      \\\hline
      {Test Cases} & 
      \begin{tabular}{m{0.03\linewidth}|@{}c@{}}
        {TC\textsubscript{03}} & 
        \begin{tabular}{m{0.08\linewidth}|m{0.64\linewidth}}
             {Modification} & {Introduction of one bird to 30\% of the training images.} \\\hline
             {Variation} & {10 different birds chosen at random for each image.}\\
        \end{tabular}
        \\\hline
        {TC\textsubscript{04}} & 
        \begin{tabular}{m{0.08\linewidth}|m{0.64\linewidth}}
             {Modification} & {Introduction of one tree to 30\% of the training images.} \\\hline
             {Variation} & {10 different trees chosen at random for each image.}\\
        \end{tabular}
        \\
        \end{tabular} \\
    \end{tabular} \\\hline
    {MR\textsubscript{04}} &
   \begin{tabular}{m{0.1\linewidth}| @{}c@{}}
      {Description} &
      \begin{tabular}{m{0.8\linewidth}}
      {Introduction of a minimally obtrusive object consistently in all training images should result in a grey-tinted effect and impact the IS in a similar manner, regardless of the colour of the object introduced.}
      \end{tabular}
      \\\hline
      {Causal Relation} &
       \begin{tabular}{m{0.8\linewidth}}
      {MR\textsubscript{02}, MR\textsubscript{03}}
      \end{tabular}
      \\\hline
      {Test Cases} & 
      \begin{tabular}{m{0.03\linewidth}|@{}c@{}}
        {TC\textsubscript{05}} & 
        \begin{tabular}{m{0.08\linewidth}|m{0.64\linewidth}}
             {Modification} & {Introduction of one bird to 100\% of the training images.} \\\hline
             {Variation} & {1 green bird for all images.}\\
        \end{tabular}
        \\\hline
        {TC\textsubscript{06}} & 
        \begin{tabular}{m{0.08\linewidth}|m{0.64\linewidth}}
             {Modification} & {Introduction of one bird to 100\% of the training images.} \\\hline
             {Variation} & {1 blue bird for all images.}\\
        \end{tabular}
        \\\hline
        {TC\textsubscript{07}} & 
        \begin{tabular}{m{0.08\linewidth}|m{0.64\linewidth}}
             {Modification} & {Introduction of one bird to 100\% of the training images.} \\\hline
             {Variation} & {1 red bird for all images.}\\
        \end{tabular}
        \\
        \end{tabular} \\
    \end{tabular} \\\hline
    {MR\textsubscript{05}} &
   \begin{tabular}{m{0.1\linewidth}| @{}c@{}}
      {Description} &
      \begin{tabular}{m{0.8\linewidth}}
      {Introduction of an unobtrusive object consistently in all training images should not result in a grey-tinted effect and not impact the IS by a huge factor.}
      \end{tabular}
      \\\hline
      {Causal Relation} &
       \begin{tabular}{m{0.8\linewidth}}
      {MR\textsubscript{02}, MR\textsubscript{03}, MR\textsubscript{04}}
      \end{tabular}
      \\\hline
      {Test Cases} & 
      \begin{tabular}{m{0.03\linewidth}|@{}c@{}}
        {TC\textsubscript{08}} & 
        \begin{tabular}{m{0.08\linewidth}|m{0.64\linewidth}}
             {Modification} & {Introduction of one bird to 100\% of the training images, without obstructing the actual bird-of-interest.} \\\hline
             {Variation} & {10 different birds chosen at random for each image.}\\
        \end{tabular}
        \\
        \end{tabular} \\
    \end{tabular} \\\hline
  \end{tabular}
\end{table*}

In order to validate \textbf{MR\textsubscript{01}}, we propose the first test case, \textbf{TC\textsubscript{01}} to look into the ability of the StackGAN-v2 model to manage the quantification of the focal object (the birds). We have found that the model utilises bounding boxes when searching for birds in an image and are prone to poor performance where multiple birds were not properly identified and only parts of their bodies were included in the bounding box (see Figure \ref{fig:bounding_box_issue}). Thus, we are interested to observe the ability of the model to correctly identify the actual bird within the image, instead of the added bird we introduced. 


It was our expectation that the resulting synthesised images would not be affected adversely as was stated in \textbf{MR\textsubscript{01}}. Visually, we expect to see an additional bird in all generated images, as it was consistently present in all training images. 


\begin{figure*}[b]
\includegraphics[width=0.9\textwidth]{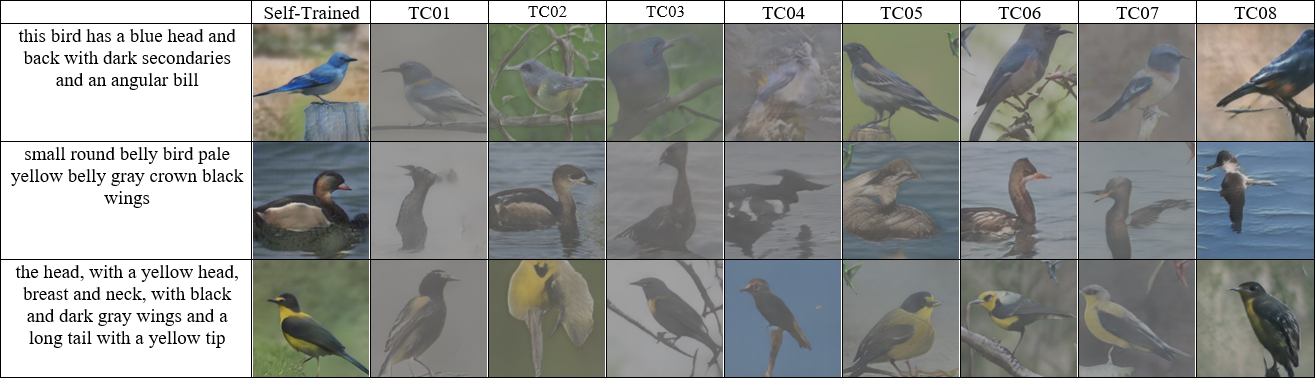}
\centering
\caption{Sample of the generated images for each test case.}
\label{fig:table_generated_images}
\end{figure*}

However, the results deviated from our expectations. Upon visual inspection, it was apparent that the resulting images were all tinted in grey colour as can be noted in Figure \ref{fig:table_generated_images}, with the colours being largely washed-out. In this regard, the quantitative analysis in Table \ref{table:1} agreed with user judgement and showed a significant decrease in both IS and Likert scores. Further examination had revealed the issue may lay solely in the third and final generator, which was the only generator to yield fake samples with a grey-tinted effect.

To elaborate, when performing training on a new model instance, multiple fake samples are generated as the output. Generally, a fake sample is the output from the generator of the StackGAN-v2 model, of which there are a total of 3. Thus, we are able to observe 3 different fake samples for a particular iteration of the training process. We observed in multiple instances that fake samples 2 were consistently tinted grey compared to the fake samples 0 and 1, which were vibrant in colour. Efforts were made to uncover the cause for such an effect, where both the StackGAN-v2 paper and source code were studied in detailed. We concluded that the colour consistency regularisation was not responsible for this effect, as was clarified in the original paper that it was not required in the text-to-image synthesis task, and thus irrelevant to the current study; the source code also revealed no distinctive difference in behaviour for the third generator compared to the other two generators. Finally, visual inspection also revealed that an assortment of behaviour was observed with the added bird in the generated images, where they may exhibit one of the following properties:
\begin{itemize}
    \item Without any of the introduced birds;
    \item With one single introduced bird;
    \item With multiple introduced birds.
\end{itemize}
This reveals an inconsistency in the StackGAN-v2 model, which may be attributed to instability or poor bounding box behaviour. The poor result generated for \textbf{MR\textsubscript{01}} indicates that the model may have mistakenly identified the added bird as the actual bird of the image. However, the model may have also been confused by the 10 different types of birds added and will thus be further investigated in other test cases. 

The results from \textbf{TC\textsubscript{01}} lead us to define two new metamorphic relations, \textbf{MR\textsubscript{02}} and \textbf{MR\textsubscript{03}}, with regards to the effect of the introduced objects on the model. \textbf{MR\textsubscript{02}} was introduced to help identify if the bounding box issue was truly prevalent in \textbf{TC\textsubscript{01}}, and if so, a different object, such as a tree should reduce the confusion of the bounding box in identifying the focal birds in the image, which implies a higher IS and better quality of the generated images. \textbf{MR\textsubscript{03}} was also introduced for a similar reason, where we assume that if a lesser extent of images is modified with extra birds, then the level of confusion introduced to the model should diminish, thus resulting in a higher IS and a less severe grey-tinted effect.

\begin{figure}[!b]
\includegraphics[width=0.45\textwidth]{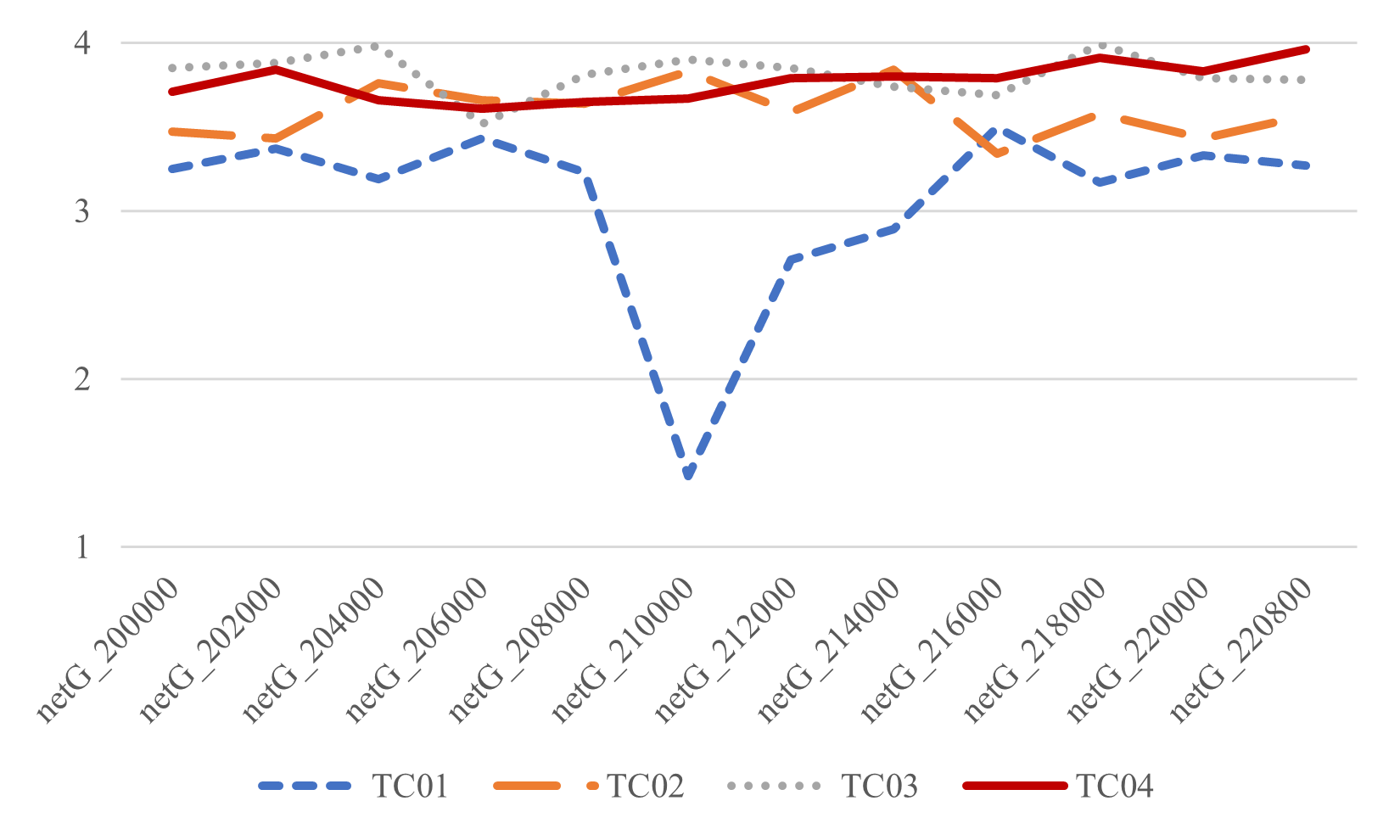}
\centering
\caption{IS for \textbf{TC\textsubscript{01}}, \textbf{TC\textsubscript{02}}, \textbf{TC\textsubscript{03}} and \textbf{TC\textsubscript{04}}. The difference between \textbf{TC\textsubscript{02}} and \textbf{TC\textsubscript{04}} is less than the other two test cases, indicating the impact of modifying a proportion of training image dataset with the bird object having a larger effect as compared to using the tree object.}
\label{fig:chart_TC1_2_3_4}
\end{figure}

The second test case, \textbf{TC\textsubscript{02}}, was built on the \textbf{MR\textsubscript{02}} 
, we expect there to be a similar effect as observed in \textbf{TC\textsubscript{01}}, especially so if the bounding box was not misguided by the newly introduced objects.
While the result was consistent with \textbf{MR\textsubscript{02}}, we observed that the grey-tinted effect was less prominent in this particular test case, and the IS also exhibited a higher value than \textbf{TC\textsubscript{01}} (see Table \ref{table:1}). This observation, where the result of \textbf{TC\textsubscript{02}} was significantly better than the \textbf{TC\textsubscript{01}}, implies that the type of object introduced does impact the model differently. This effect can be attributed to the different shape, colour, position, or other variables of the two test cases and will be further explored in the subsequent test cases.

Following this, the third test case \textbf{TC\textsubscript{03}} building on the \textbf{MR\textsubscript{03}}, was conducted to observe the stability of the StackGAN-v2 model with respect to the proportion of the modified training image dataset.  
Consistent with \textbf{MR\textsubscript{03}}, it was observed that the grey-tinted effect was much less noticeable in \textbf{TC\textsubscript{03}} compared to its predecessor. The IS also reveals the highest value amongst all the test cases, substantiating the claim from \textbf{MR\textsubscript{03}}. 

At this stage of our experiments, we are interested to verify that the \textbf{MR\textsubscript{03}} is also applicable to the test case when the foreign object is a tree.
The results of the \textbf{TC\textsubscript{04}} were found to be faithful to \textbf{MR\textsubscript{03}}, similarly to its predecessor, \textbf{TC\textsubscript{03}}. Despite this, we were able to interpret a few irregularities in the results of Table \ref{table:1}. It had been earlier remarked that the \textbf{TC\textsubscript{02}} had a significantly higher IS when compared to \textbf{TC\textsubscript{01}}. Thus, we have assumed that when the added object is a tree, we  may expect it to have a higher IS than a similar scenario where the added object is a bird. In contradiction to this assumption, the results from \textbf{TC\textsubscript{03}} were found to produce a higher IS than \textbf{TC\textsubscript{04}}. Interested to pursue this irregularity further, we began checking if this was just an isolated case and not a consistent behaviour. This lead to the finding that the effect of modifying only 30\%, rather than 100\% of the training image dataset for trees is minute compared to the effect in birds as can be observed from Figure \ref{fig:chart_TC1_2_3_4}. The observation was also in contradiction with \textbf{TC\textsubscript{02}}, where it was assumed that the impact of the grey-tinted effect and IS would be consistent for all test cases regardless of the type of object. This presents multiple possible scenarios as the cause for such an irregularity, of which we have established two:
\begin{itemize}
    \item One such theory hypothesise that colour may be a factor in attributing the grey-tinted effect;
    \item The second theory is built on the idea of shape of the added object being a contributing factor to the grey-tinted effect in the generated images. Theoretically speaking, the effect would be especially prominent if the object was a bird, since it adds more confusion to the bounding box in trying to identify the focal bird in the image.
\end{itemize}

\begin{figure}[!b]
\includegraphics[width=0.45\textwidth]{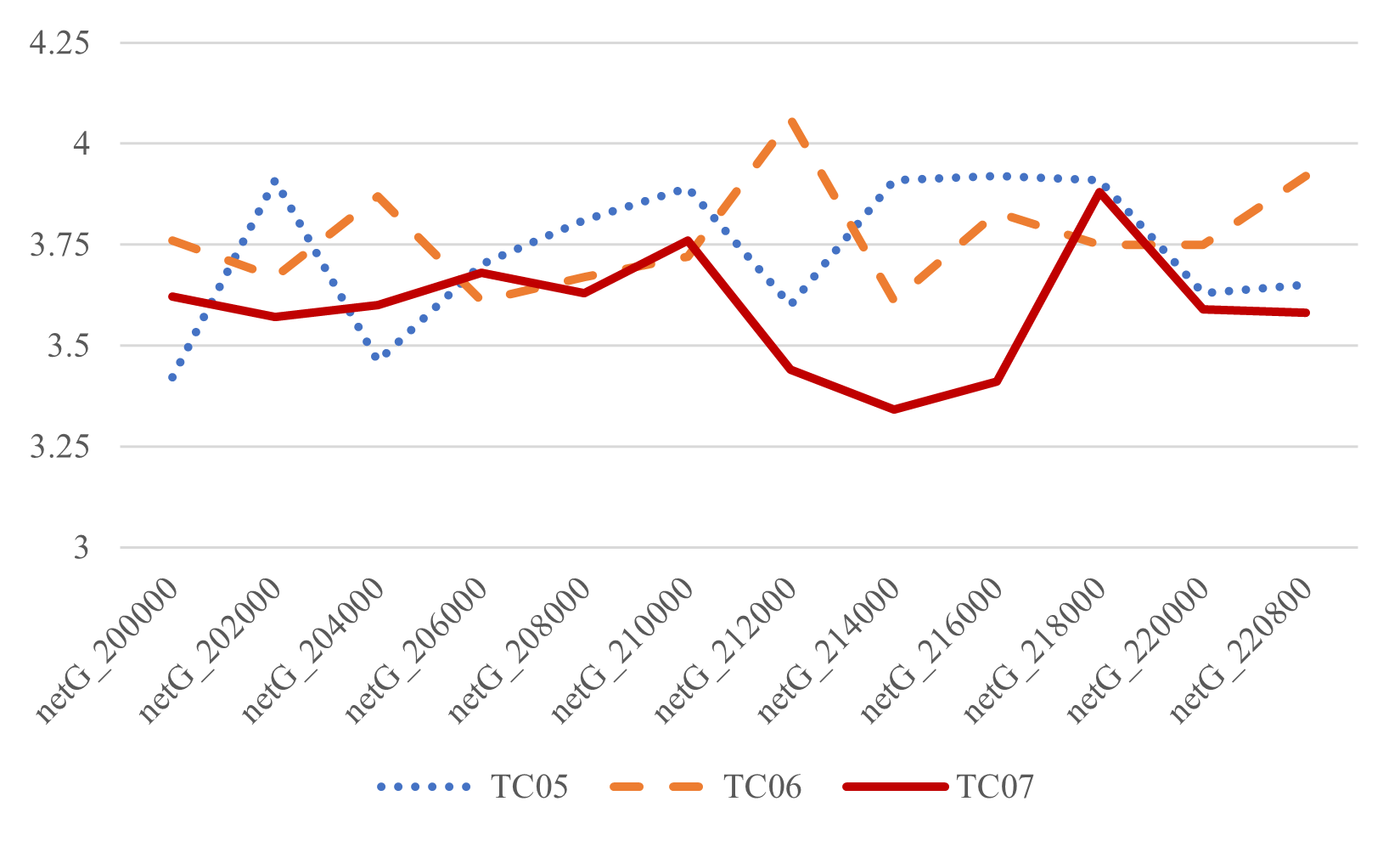}
\centering
\caption{IS for \textbf{TC\textsubscript{06}}, \textbf{TC\textsubscript{07}}, and \textbf{TC\textsubscript{08}}. Noted that \textbf{TC\textsubscript{08}} in particular achieved the worst IS while \textbf{TC\textsubscript{07}} achieved the best IS. This may be an indication that the colours of the introduced object have a significant effect on the StackGAN-v2 model. }
\label{fig:chart_TC5_6_7}
\end{figure}

We then proposed our fourth metamorphic relation \textbf{MR\textsubscript{04}} and its corresponding test cases \textbf{TC\textsubscript{05}}, \textbf{TC\textsubscript{06}} and \textbf{TC\textsubscript{07}}, based on the first theory identified from the result of \textbf{TC\textsubscript{04}} where the colour of the added object may or may not be an attributing factor in the grey-tinted effect.
The test cases \textbf{TC\textsubscript{05}}, \textbf{TC\textsubscript{06}} and \textbf{TC\textsubscript{07}} were designed such that we may review if specific colours in the RGB scheme may prove a difficult concept for the StackGAN-v2 model to grasp and visualise. In particular, we have reason to believe that the model may struggle explicitly with the colour green and/or brown, based on our observations with the previous test cases \textbf{TC\textsubscript{03}} and \textbf{TC\textsubscript{04}} concerning trees. In order to ensure the consistency and reliability of our results, the test cases \textbf{TC\textsubscript{06}}, \textbf{TC\textsubscript{06}}, and \textbf{TC\textsubscript{07}} were implemented using the same bird image but represented in the respective colours specified by the test cases, and are all positioned identically for a particular image in each test case.

The results of \textbf{TC\textsubscript{05}} conformed to our expectations, where the grey-tinted effect was still visibly present. The IS for \textbf{TC\textsubscript{06}}, \textbf{TC\textsubscript{06}}, and \textbf{TC\textsubscript{07}} was placed between \textbf{TC\textsubscript{01}} and \textbf{TC\textsubscript{03}}, which was expected, considering the fact that \textbf{TC\textsubscript{05}} had introduced more noise to the sample compared to \textbf{TC\textsubscript{03}}, where only 30\% of the images were modified and thus introduce less confusion to the StackGAN-v2 model (see Table \ref{table:1}). Both \textbf{TC\textsubscript{06}} and \textbf{TC\textsubscript{07}} displayed consistent attributes to what had been described in \textbf{TC\textsubscript{05}}, with the exception of \textbf{TC\textsubscript{06}}, where we notice that it has a higher IS as compared to \textbf{TC\textsubscript{03}}. This is unexpected since \textbf{TC\textsubscript{03}} had a much smaller proportion of images (30\% as compared to 100\%) with added objects, which in accordance to \textbf{MR\textsubscript{03}}, was anticipated to result in a higher IS. In order to identify the cause for this irregularity, we began looking into the differences between \textbf{TC\textsubscript{03}} and \textbf{TC\textsubscript{06}} in closer detail:
\begin{itemize}
    \item Proportion of modified images where \textbf{TC\textsubscript{03}} had 30\% and \textbf{TC\textsubscript{06}} had 100\%;
    \item \textbf{TC\textsubscript{03}} randomly chooses 1 of 10 different birds to add to a training image, while \textbf{TC\textsubscript{06}} only introduces a single type of bird in blue colour. While it may appear that both colour and shape could be a contributing factor here, 
    it seems likely that shape could be overruled as a contributing factor given that 
    the results from \textbf{TC\textsubscript{05}} and \textbf{TC\textsubscript{07}} had a lower IS than \textbf{TC\textsubscript{03}};
    \item Positions in which birds were added to the images were not consistent. 
    Further adjustments were necessary to ensure minimal obstruction of the focal bird in the image.
\end{itemize}
From our results, the most feasible reason for the observed irregularity appears to be either the colour or the position of the birds. Upon closer inspection, we believe colour has little to no effect on the model as the 3 test cases \textbf{TC\textsubscript{05}}, \textbf{TC\textsubscript{06}}, and \textbf{TC\textsubscript{07}} had relatively similar results. 

Having now established that both shape and colour are unlikely to be the contributing factor, we began evaluating how the different positions of the birds could be affecting the results in such a manner. As was emphasised in all our test cases, we make an effort to \textit{minimise} the obstruction of the added object on the focal birds. In hindsight, while we are able to guarantee that the focal birds are not being absurdly overshadowed by the newly added objects, there may exist instances where the added object lightly brushes the focal bird's wings or torso.
This became an even more relevant topic when we consider the previously examined bounding box issue. We now hypothesise that if a certain proportion of the birds in the training images is obscured by the object in some form, it would affect the ability of the bounding box to correctly identify the focal bird within the image, which is applicable to the irregularity in \textbf{TC\textsubscript{06}} if it had only a very small proportion of mistakenly identified bounding boxes when compared to \textbf{TC\textsubscript{03}}. To simplify, while \textbf{TC\textsubscript{06}} had a much larger proportion of modified images, it is possible that the bounding box was able to identify the focal birds more accurately in this particular test case if the introduced birds did not obscure the focal bird in any way. We now define a new metamorphic relation \textbf{MR\textsubscript{05}} and its resulting test case \textbf{TC\textsubscript{08}} to further investigate this hypothesis. 
In \textbf{TC\textsubscript{08}}, we hope to compare its results with \textbf{TC\textsubscript{01}} to attest our claim that the obstruction of the view of the focal bird would impede the ability of the bounding box to correctly identify it. The resulting model substantiates our claim with \textbf{MR\textsubscript{05}}, exhibiting a much higher IS than \textbf{TC\textsubscript{01}}, as can be noted from Table \ref{table:1}. Another surprising detail was observed, where we found the generated images to be completely free of any grey-tinted effect (see Figure \ref{fig:table_generated_images}). This further supports our claim that the bounding box is easily confused if an object obscures the focal body of the image, and is likely the contributing factor to the grey-tinted effect.

Finally, we review the results of our Likert Scale analysis for all test cases (see Table \ref{table:1}). Generally, we found that the IS model corresponded well with human judgement in our analysis, with the exception of \textbf{TC\textsubscript{04}}, where it was found to be represented most poorly in both the semantic and realistic senses. This is in contradiction with the results from the IS, and may act as a point of reference in future investigations.

\begin{figure}[t] 
\includegraphics[width=0.4\textwidth]{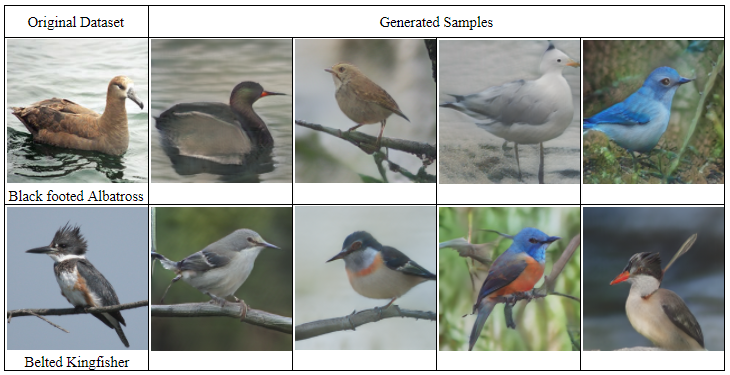}
\centering
\caption{Sample of generated images from the self-trained model. }
\label{fig:inconsistent_species}
\end{figure}

\begin{figure}[t] 
\includegraphics[width=0.3\textwidth]{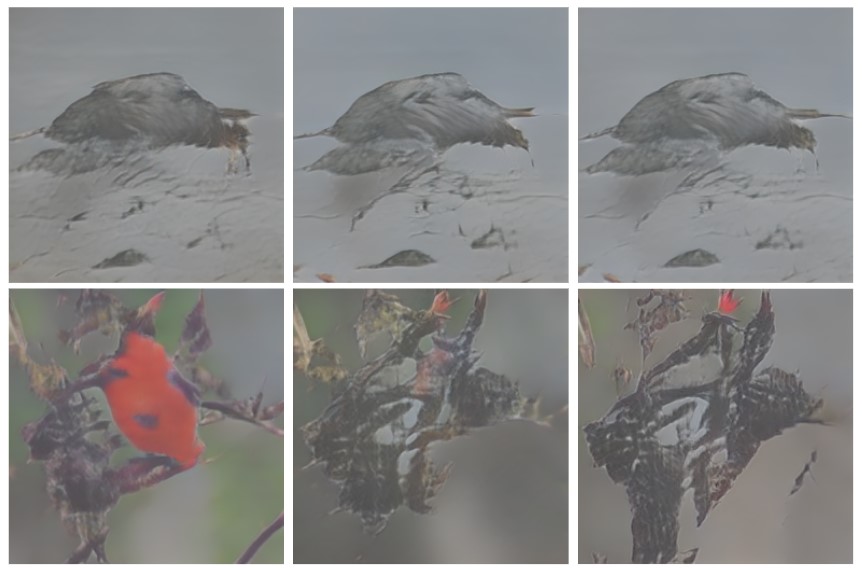}
\centering
\caption{Sample of generated images from \textbf{TC\textsubscript{05}}. The deformed images seem to follow some form of pattern, thus appearing repetitive in nature.}
\label{fig:deformed_birds}
\end{figure}

When simply observed in a qualitative manner, we believe some anomalies appeared consistently in all test cases:
\begin{itemize}
    \item The ability of the model to differentiate between species of birds are relatively weak. Different bird species were sparsely distributed in each other's categories, even in the self-trained model instance (see Figure \ref{fig:inconsistent_species});
    \item The model has a tendency to generate multiple similar deformed images in a single test case (see Figure \ref{fig:deformed_birds}).
\end{itemize}

\section{Conclusion}
In this study, we proposed a metamorphic testing approach to identifying the robustness of text-to-image synthesis models, specifically the StackGAN-v2 model. The first contribution of this study is the identification of five metamorphic relations that may be applied to any text-to-image synthesis model to verify the robustness of the algorithm in the presence of unexpected input training images. The metamorphic relations were derived successively based on one naive metamorphic relation as a starting point, which is applicable to any text-to-image synthesis model : \emph{``How would the text-to-image synthesis model behave when noises were introduced to the training images?''}. Our findings suggest that by using the prior test case results, it is possible to derive more complicated metamorphic relations and test cases to evaluate the robustness of text-to-image synthesis models. This can then follow an iterative or cyclic process model, instead of traditional metamorphic testing methods which tend to follow a waterfall process model. Our subsequent contribution is the identification of the robustness issues in the StackGAN-v2 model, which we have found to be subjected to undocumented behaviour, particularly the grey-tinted effect when unexpected input training images are utilised. With the aid of metamorphic testing, we have successfully revealed the aforementioned behaviour of the StackGAN-v2 model, which are otherwise hidden from the researchers. Ultimately, our findings can help researchers to better understand the behaviour of the StackGAN-v2 model and further develop a robust image synthesis model.

There are some potential directions to approach for future research. The naturalness of the modified images from the dataset can be enhanced since we think that realism and naturalness will not significantly affect the result from metamorphic testing. Improve the naturalness of the modified training images may benefit this research by revealing the errors which are more likely found in real life. Other than that, another limitation of this work is that we believe it is limited in its characteristic of requiring intensive labour for qualitative analysis. As such, we intend to automate our metamorphic testing process in future research. In addition to that, we are also interested in looking into the following topics: 1.) to perform further examination of the impact of colour in the StackGAN-v2 model; 2.) to investigate the effects of shape of the added object on the StackGAN-v2 model; 3.) to incorporate text input modifications in metamorphic testing.


\section*{Acknowledgment}
We would like to thank all the authors of the StackGAN-v2 paper \cite{zhang2018stackgan++} 
for responding to our queries. This work was partly supported by the Advanced Engineering Platform's CyberSecurity \& AI $\varphi^{2}$ Cluster Funding (AEP-2020-Cluster-04), Monash University, Malaysia campus.



%
\printbibliography
\end{document}